\documentclass[lettersize,journal]{IEEEtran}
\usepackage{amsmath,amsfonts}
\usepackage{amssymb}
\usepackage{algorithmic}
\usepackage{algorithm}
\usepackage{array}
\usepackage[caption=false,font=normalsize,labelfont=sf,textfont=sf]{subfig}
\usepackage{textcomp}
\usepackage{comment}
\usepackage{float}
\usepackage{stfloats}
\usepackage{url}
\usepackage{verbatim}
\usepackage{graphicx}
\usepackage{color}
\usepackage{cite}
\newcommand{\mmax}[1]{{{\color{black}#1}}}

\begin{document}

\title{Thermopneumatic Pixels: Fast, Localized, \\ Robust, Low-Voltage Touch Feedback}
%

\author{Max Linnander \IEEEmembership{Student Member,~IEEE}, Yon Visell~\IEEEmembership{Member,~IEEE,}
\thanks{M. Linnander, and Y. Visell are with the Department of Mechanical Engineering, University of California, Santa Barbara, 93106 USA (e-mail: maxlinnander@ucsb.edu; yonvisell@ucsb.edu).} 

\thanks{Manuscript received Month XX, XXXX; revised Month XX, XXXX.}
}



\maketitle

\begin{abstract}
We present thermopneumatic pixels (TPPs) -- low-profile pixels and arrays that generate dynamic tactile feedback.  These devices are thin, fast, reconfigurable, and output localized transient displacements at each pixel.  Their parsimonious design -- a layered architecture without internal moving parts -- and low-voltage ($\lesssim$10 V) operation may facilitate practical integration in a wide variety of interfaces. Each TPP converts brief electrical pulses into transient air pressure increases in an internal cavity, yielding out-of-plane forces and displacements for tactile feedback. We demonstrate TPPs that output displacements of 1 mm and forces exceeding 1 N, with millisecond response times, in packages that are less than 3 mm thick. Force and displacement increase with pixel surface area, facilitating tailorability.  The pixels can also generate oscillating feedback at pulse rates up to 300 Hz range.   We report designs for compact arrays of pixels at 4 mm spacing, and simple pulse driving architectures using miniature transistors driven by microcontrollers. We characterize the mechanical, dynamic, and thermal response of TPPs, and their robustness and consistency over tens of thousands of cycles.  We report perceptual experiments on spatial localization and intensity as a function of driving power.  Together, these results establish thermopneumatic pixels as a compact, adaptable tactile technology that blends performance and practicality. 




\end{abstract}


\section{Introduction}
\begin{figure*}[t]
\begin{center}
\includegraphics[width = 182mm]{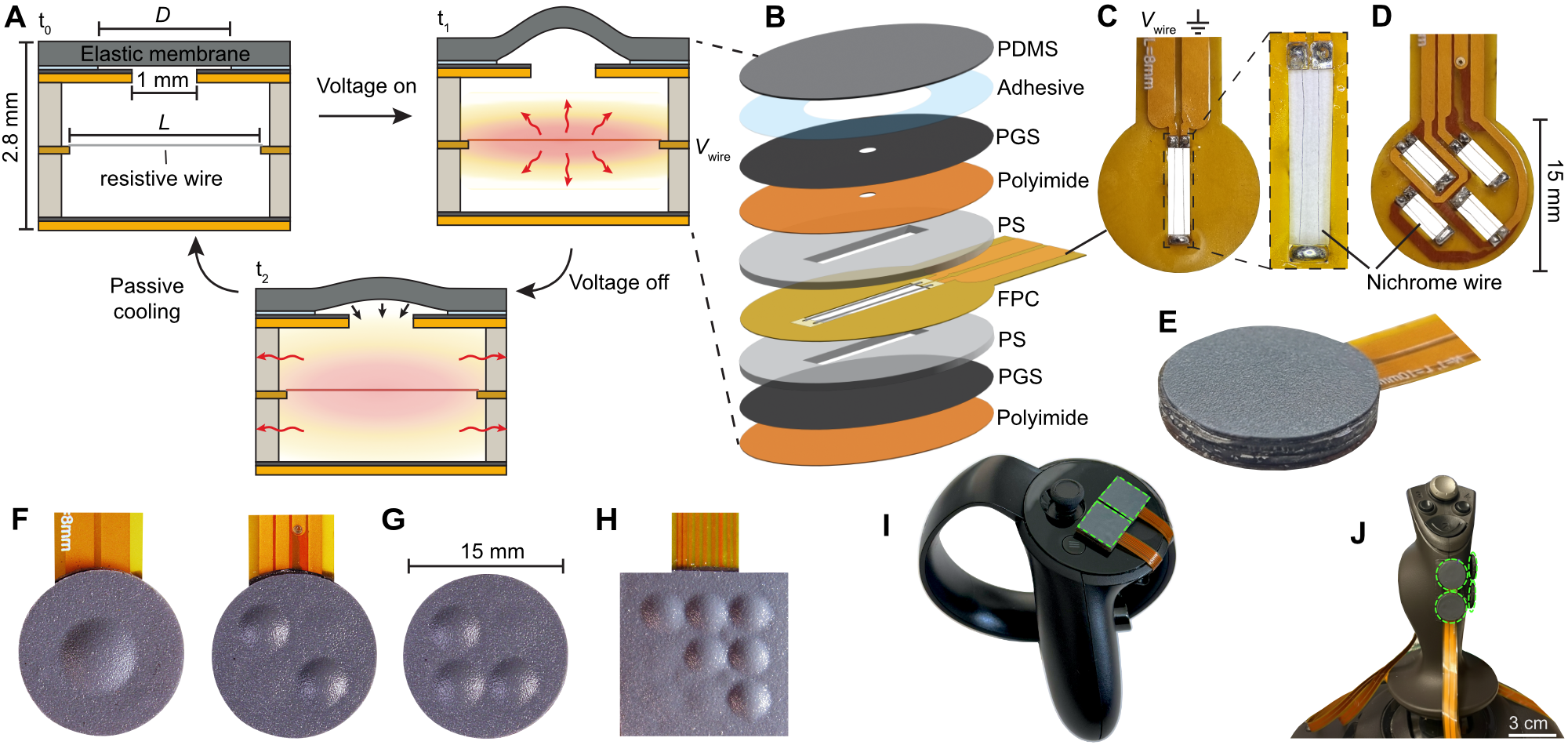}
\end{center}
\caption{A) Operating principle for thermopneumatic pixels (TPPs). TPPs deliver heat to air encapsulated within a small sealed cavity, driving gas expansion that yields localized forces and displacements.  Heat is supplied to the gas (air) via transient Joule heating of a resistive Nickel-Chromium (NiCr) wire suspended in the cavity.  Due to the small diameter of the wire (48 AWG), heat is rapidly transferred to the gas, driving a rapid pressure increase in the gas.  The air pressure increase drives deflection of an elastic membrane sealing the top of the cavity, yielding localized forces and displacements. B) The design uses a simple, layered assembly of common materials.  Exploded view of the multilayer TPP stack, comprising patterned structural, thermal, and compliant layers. C) Photograph of flexible printed circuit (FPC) layers with suspended NiCr wires, for a module with a single small TPP (cavity length $L = 8$~mm). The dashed box shows an enlarged view of the NiCr wire. D) FPC  layer for a quartet module with four TPPs ($L = 4$~mm). E) Photograph of the module. F) Top-down view of TPP during actuation ($L=8$~mm, pixel diameter $D=6$~mm) G) Top-down view of the quartet module, with four TPPs, as it displays different patterns ($L=4$~mm, $D=4$~mm).
\mmax{H) Top-down view of a 9-TPP module, as it displays a pattern ($L=4$~mm, $D=3$~mm). I, J) Examples of how TPP arrays can be integrated in controllers.  Green dashed squares: 3$\times$3  arrays; circles: 2$\times$2 arrays.}}
\label{fig:concept}
\end{figure*}

Haptic technologies have the potential to endow computational systems with physical expression and agency in ways that are matched to the human sense of touch. Over the past half century, haptics research has produced numerous  technologies capable of rendering tactile feedback with far greater specificity than the  vibrotactile devices that dominate most commercial products.  However, the translation of these technologies into deployed interfaces or products has remained extremely limited.  The key barrier is not only performance, but also the difficulty of achieving high fidelity, localized, precise tactile output in forms that are compatible with practical constraints arising from manufacturability, dimensions, packaging,  material compatibility, and electronic interfacing requirements are paramount.  As a result, the great majority of haptic technologies remain confined to the laboratory, irrespective of the performance they achieve.  High-fidelity tactile devices that are manufacturable and suitable for product integration and validation  could greatly expand the role of haptics in numerous computing interfaces and products, including wearables, writing instruments, vehicle controls, medical instruments, and other systems \cite{wang2019multimodal,lee2004stylus,colgate2013haptic, okamura2009medical}.

In assessing requirements for such technologies, it is valuable to bear in mind that it is impractical today, or for the forseeable future, to realize an idealized haptic technology that matches the  capacities of human sensation and perception \cite{biswas2019materials}.  Thus, the performance and fidelity requirements for tactile technologies are often determined based on criteria that are strongly application dependent.    Force, displacement, and temporal performance characteristics that support perceptual fidelity vary with several application-specific factors, including anatomical site, contact geometry, loading conditions, and feedback requirements \cite{jones2006human}.  
Nonetheless, the utility of any tactile technology, and its potential breadth of application, is enhanced to the extent that it is capable of a wide operating gamut matched to the sensing capabilities of the human skin.  
For tactile (light touch) devices, desirable characteristics include the ability to furnish feedback that can be localized within a few millimeters, to produce responses with temporal precision on the order of tens of milliseconds (supporting persistence of perception across actuation cycles), to generate surface deformations on the order of millimeters or forces (driving deflections of the skin)  or hundreds of millinewtons \cite{copeland2010identification,craig1987vibrotactile}.



\mmax{Many actuator technologies are able to approximate such objectives, but practical implementation often exposes tradeoffs that can constrain replication and deployment in compact systems. Electromagnetic devices serve the majority of needs in actuation \cite{choi2012vibrotactile,kern2023engineering}.  However, as integration density increases, significant challenges manifest, such as the following: constraints and overhead from packaging of power inductors, magnetic shielding requirements, and heat transfer and thermal scaling effects that are specific to electromagnetic devices  \cite{kastor2023ferrofluidic,streque2012emactuator,strasnick2016magneticactuators}. Electrostatic dielectric-elastomer actuators, and their variants, can achieve high bandwidth and substantial strain \cite{leroy2020haxel,zhao2018dea,ji2021dea}, but demand drive voltages of multiple kilovolts that in turn raise issues \cite{leroy2023haxel,grasso2023haxel,zhao2018dea,mitchell2022HV}. Other emerging approaches, including electroactive polymers \cite{frediani2021eap,koo2008soft,qiu2018polymerdisplay}, electroosmotic and fluidic devices \cite{shultz2023flatpanel,shen2023fluidreality,shan2024hydraulic}, phase-change systems \cite{ueno2020phasechange,uramune2022hapouch}, and soft pneumatic architectures that utilize external pressure sources  \cite{russomanno2015pneumatic,russomanno2017pneumatic,lining2013pneumatic}, expand the available design space and can offer attractive combinations of compliance, form factor, and spatial addressability. At the same time, these technologies often introduce embodiment-specific compromises in drive complexity, encapsulation, response speed, lifetime, or fabrication process compatibility \cite{biswas2019materials,kato2007polymeractuator,shan2024hydraulic,wu2012pneumatic}. Their continued development is promising, but the breadth of their practical utility in compact, manufacturable haptic products is still being established.

Thermal actuation mechanisms can offer several advantages, even if traditional architectures exhibit well-known drawbacks. Thermal devices often require lower complexity mechanical assemblies than electromechanical devices, can be operated with simple electronic drive circuits, and can be operated at low voltage \cite{biswas2019materials}\cite{hwang2021bimorph}\cite{qiu2018polymerdisplay}. However, many thermally driven mechanisms are fundamentally constrained by thermal inertia and heat transport. Most such devices involve the deposition of heat in solid or liquid masses, yielding slow response times \cite{camargo2012opticalbraille, torras2014optomechanical,besse2017fea,hiraki2020laserpouch,uramune2022hapouch}.  Fundamentally, the slow responses exhibited by such materials are due to magnitude of the heat transfer timescale, $\tau$, which is proportional to the volume specific heat capacity, $\rho c_p$, where $\rho$ is mass density and $c_p$ is specific heat capacity.  For solid and liquid bodies,  $\rho c_p \sim~$1 J/cm$^3$K.  Thus, in  actuators involving heat transfer to such materials, response times are often in the $10^{0}$–$10^{2}$~s range (i.e. seconds to hundreds of seconds) \cite{camargo2012opticalbraille, torras2014optomechanical,besse2017fea,hiraki2020laserpouch,uramune2022hapouch}, which limits their utility in haptics.}


Thermopneumatic approaches offer the potential for faster responses by heating a captive gas to perform mechanical work. Gases have volumetric heat capacities on the order of $\rho c_p \sim$0.001~J/cm$^3$K, which is orders of magnitude smaller than solids and liquids ($\sim$0.001~J/cm$^3$K), facilitating far more rapid responses. Although such configurations have been studied, the heating element has often been integrated in the wall of the gas-filled cavity, yielding long response times and low forces due to heat loss to the walls \cite{mazzotta2023thermopneumatic,mazzotta2025thermopneumatic}. Photo-thermopneumatic actuators have circumvented this by depositing radiative heat onto a low-thermal-inertia absorber, suspended in the working gas cavity. This has enabled rapid transfer of heat into the gas, but requires additional optical components and losses due to the conversion of electrical into optical energy \cite{linnander2025tactile,linnander2026hled}. These comparisons motivate our concept of suspending a thin resistive wire in a cavity encapsulating a captive gas.

In this paper, we introduce thermopneumatic pixels (TPPs), a family of tactile devices that are practical, thin, fast, reconfigurable, and yield spatially localized displacement at each pixel.  They may be suitable for integration within a wide variety of haptic systems, via graspable interfaces, interactive surfaces, wearables, or handheld devices.  Several design features make them very amenable to integration.  They can be tailored, in dimensions or output, to application requirements,  can be manufactured using standard processes, and driven at low voltages.  The lack of internal moving parts reduces packaging requirements and supports robustness.  These properties make the TPPs compatible with recent developments in software for rapid haptic design and authoring, with emerging modular architectures, and application-specific middleware that together have the potential to broaden the use of tactile technologies in numerous domains \cite{goetz2023dynamic,tummala2024skinsource,huang2025vibraforge,rokhmanova2025platform}.

Each TPP consists of a sealed cavity containing a suspended resistive wire that is electrically heated using low-voltage drive signals ($\sim$10~V). Due to its low thermal inertia, the wire rapidly heats, transferring energy to the encapsulated gas, which produces a transient increase in pressure that deflects an elastic membrane and generates out-of-plane force and displacement. This mechanism enables forces exceeding 1~N and displacements approaching 1~mm with response times on the order of 5--100~ms, satisfying key requirements for high-fidelity tactile stimulation. We further present driving electronics capable of operating multiple modules concurrently, illustrating the modularity and ease of deployment of the system. Finally, we characterize the mechanical, dynamic, and thermal behavior of the TPPs and report perceptual results demonstrating robust, perceivable tactile feedback across a broad range of drive conditions.

\section{Design and Operating Principle}

TPPs deliver fast, localized tactile feedback by heating air encapsulated in small volumes (6 to 32 $\mu$L) in order to drive rapid pressure increases that deflect a top membrane, forming the pixel's touch surface.  Their fast responses are facilitated by a favorable physical scaling regime, notably the high rate of heat transfer that is achievable at small dimensions, and the low thermal mass of air. 

Each TPP operates on an electro-thermo-mechanical principle, in which electrical energy is converted into heat within a thin resistive wire when a voltage, $V_\mathrm{wire}$, is applied across it (Fig.~\ref{fig:concept}A). Owing to the low thermal mass of the wire, its temperature rises rapidly, producing a transient increase in the temperature and pressure of the enclosed gas. The resulting gas expansion deflects an elastic membrane, generating dynamic forces $F(t)$ and displacements $z(t)$. Following termination of the electrical pulse, heat dissipates to the surrounding structure and environment, allowing the TPP to return to its initial state. Due to the physics governing heat transfer, TPPs are designed to be driven by brief energetic electrical pulses, and trains of such pulses.  Driving them in this fashion avoids unnecessary energetic losses, and produces rapid mechanical forces and displacements that are highly salient to touch.



This actuation strategy admits many possible implementations; here we describe one realization. To support scalable fabrication and ease of integration, we employed a simple, layered architecture (Fig.~\ref{fig:concept}B). Joule heating is supplied via an active layer, comprising a flexible printed circuit (FPC, 120~$\mu$m thick) with a rectangular cutout of length $L$ and width $w$.  For the TPPs demonstrated here, $L$ ranges from 2 to 10~mm, and is held constant at $w=$2~mm.  Two fine, 48~AWG Nickel-Chromium (NiCr) wires are suspended along the length of the cutout (Fig.~\ref{fig:concept}C). The wires are electrically connected at one end of the cavity to form a continuous resistive heating element.

On both sides of the FPC are vinyl-cut and laser-cut layers of polyimide (125~$\mu$m), pyrolytic graphite sheet (PGS, 17~$\mu$m), polysiloxane (PS, 0.8~mm), adhesive, and a compliant polydimethylsiloxane (PDMS) membrane (250~$\mu$m) that functions as the tactile surface. The PS layers contain  $L \times w$~mm cutouts aligned with the FPC aperture.  Together these form the cavity. Narrow apertures in the overlying polyimide and PGS layers create a central opening that permit pressurization of the cavity to deflect the PDMS membrane. The effective pixel diameter $D$, i.e. the diameter of the outward expansion, determined by the circular cutout in the adhesive layer beneath the PDMS (Fig.~\ref{fig:concept}A). This parameter may be adjusted.  Other geometries can also be used. Multiple or many resistive heating wires can be used.   A full list of materials can be found in Appendix \ref{appx:tables}. 

The effective gap distance from the resistive wire(s) to the cavity walls, i.e. the effective air gap size, determines how long heat can be supplied to the heating element before losses in the wire-air-wall heat transfer pathway become pronounced.

We use NiCr as the heating element due to its thermal stability and resistance to oxidation at elevated temperatures, and incorporate pyrolytic graphite sheets (PGS) to provide high in-plane thermal conductivity for passive lateral heat spreading. To support rapid prototyping and straightforward integration into interactive systems, we implement modules as slim, mechanically flexible circular pads (15~mm diameter, 2.8~mm thick) that can be mounted on curved or planar surfaces and configured with one or multiple TPPs (Fig.~\ref{fig:concept}E--J).  

\begin{figure}
\begin{center}
\includegraphics[width = 85 mm]{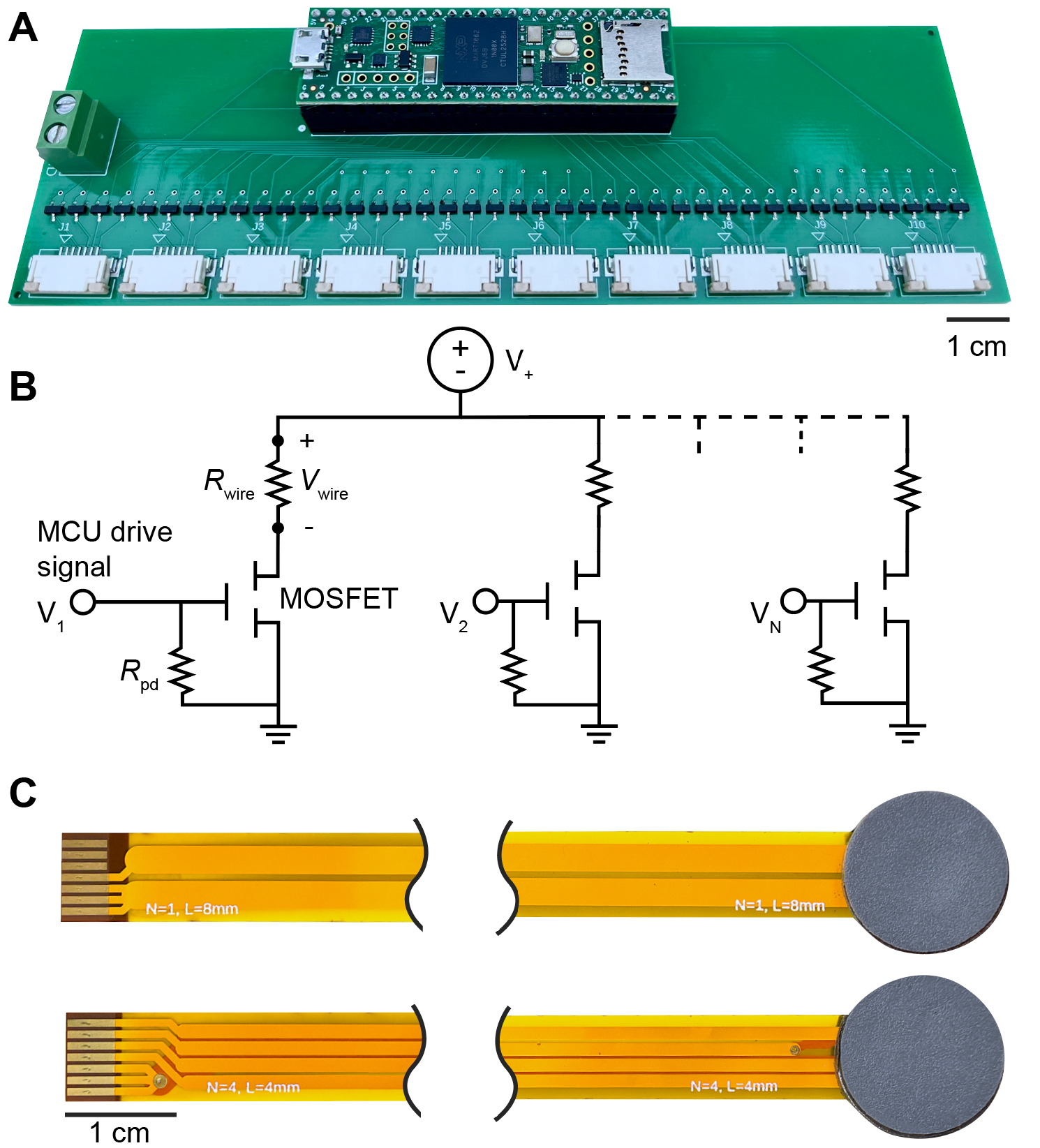}
\end{center}
\caption{A) Electronic driver board interfacing with up to ten modules, enabling simultaneous operation of up to forty TPPs in the quartet configuration. B) General schematic used to drive the TPPs. C) The single and quartet modules share a common electrical pinout to facilitate operation with the driver board. The photos show modules with a single TPP (top, $L = 8$~mm) and quartet (bottom, $L = 4$~mm). The connector length is application-dependent.  
}
\label{fig:electric}
\end{figure}

To demonstrate the ease with which these modules can be driven, we designed an electrical driver board (Fig.~\ref{fig:electric}). We implemented a common electrical pinout for driving both configurations shown in (Fig.~\ref{fig:concept}F--G) -- the quartet of four smaller TPPs and the single TPP. The electronic driver board is capable of operating up to ten such modules, enabling simultaneous operation of up to forty TPPs in the quartet configuration or ten in the single TPP configuration shown. The modules connect to the driver board via zero-insertion-force connectors (Fig. \ref{fig:electric}C). Each TPP is driven by brief, voltage-controlled electrical pulses (0.5–100 ms) supplied by an external programmable power supply and gated by a MOSFET switch. Pulse timing is generated directly on the microcontroller unit (MCU), which interfaces with a host computer and delivers gate-drive signals to the MOSFETs. Each gate is connected to a pull-down resistor, $R_\mathrm{pd}$, to prevent floating when inactive (Fig. \ref{fig:electric}B). Additional system-level details are provided in Appendix \ref{appx:tables}.


\section{Actuator Characterization}
\subsection{Thermomechanical Characterization}
\begin{figure*}
\begin{center}
\includegraphics[width = 182mm]{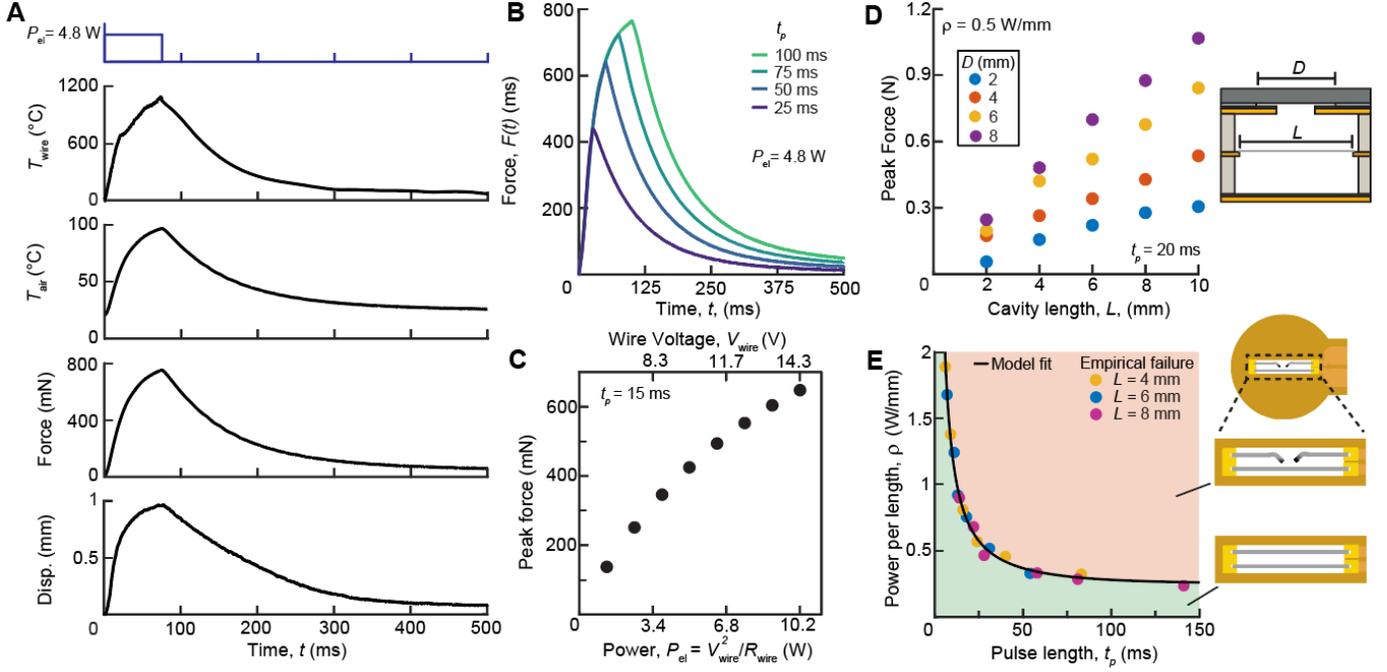}
\end{center}
\caption{Thermomechanical characterization of TPPs. A) NiCr wire temperature $T_\mathrm{wire}(t)$, inferred cavity air temperature $T_\mathrm{air}(t)$, isometric force $F(t)$, and free displacement $z(t)$ in response to a 75-ms, 4.8~W electrical pulse. 
B) Force response $F(t)$ for increasing pulse durations $t_p$ at fixed electrical power. 
C) Peak force as a function of electrical power dissipated in the wire, $P_\mathrm{el} = V_\mathrm{wire}^2/R\mathrm{wire}$, for $t_p = 15$ ms. D) Peak force measured across actuators with varying cavity length $L$ and aperture diameter $D$, at fixed power per unit length $\rho = 0.5$ W/mm. 
E) Operating envelope of TPPs, defined by power per unit length $\rho$ and pulse duration $t_p$. Shaded regions indicate viable operation (green) and empirical failure (red). Points denote measured failure events for varying $L$, and the black curve shows a fit to an analytic thermal model. 
}
\label{fig:mechanical}
\end{figure*}


We characterized the thermal and mechanical responses of the actuator under electrical pulse excitation with duration $t_p$ and voltage $V_\mathrm{wire}$ (Fig.~\ref{fig:mechanical}). The force output at the tactile membrane, $F(t)$, was measured isometrically, using a load cell (Futek LSB200).  Free (unloaded) displacement was measured using a laser triangulation sensor (Keyence IL-065). The cavity air temperature, $T_\mathrm{air}(t)$, was computed as follows.  From the measured force $F(t)$, we computed the gauge air pressure $P(t) = F(t) / A$, where $A = \pi (D/2)^2$ is the pixel area, where $D$ is the membrane aperture diameter. Air temperature was obtained using the ideal gas law, $T_\mathrm{air}(t)=T_0\!\left(F(t)/(P_0\, A)+1\right)$, where $P_0$ and $T_0$ are the initial cavity pressure and temperature. The temperature of the NiCr heating element, $T_\mathrm{wire}(t)$, was computed from in situ resistivity measurements using a resistivity--temperature relationship (Appendix~\ref{appx:wire_temp}).

We observed the temperature of the NiCr wire, $T_\mathrm{wire}(t)$, to increase monotonically upon application of a voltage, driving corresponding increases in cavity air temperature and pressure. The resulting membrane force, $F(t)$, produced a displacement, $z(t)$ (Fig.~\ref{fig:mechanical}A). Temperature, pressure, force, and displacement exhibited closely matched temporal profiles and reached peak values near the end of the applied pulse. Over the timescales examined, these variables were strongly correlated and appeared primarily related through approximately constant gain factors, rather than distinct dynamics, consistent with prior observations in thermopneumatic actuators of similar dimensions \cite{linnander2025tactile,linnander2026hled}.

Heat transfer from the suspended wire to the surrounding air and structure can be approximated using a first-order lumped thermal model, yielding a wire temperature response of the form
\begin{equation}
T_\mathrm{wire}(t) = P_\mathrm{el} R_\mathrm{thermal}\!\left(1 - e^{-t/\tau}\right),
\label{eq:T_wire}
\end{equation}
where $P_\mathrm{el} = V_\mathrm{wire}^2/R_\mathrm{wire}$ is the electrical power dissipated in the NiCr wire, with $V_\mathrm{wire}$ and $R_\mathrm{wire}$ being the voltage and resistance across the wire. The effective thermal resistance, $R_\mathrm{thermal}$, accounts for heat transfer through the air and wire, while the characteristic time constant $\tau = R_\mathrm{thermal} C$, governs the heating and cooling dynamics, where $C$ is the heat capacity of the wire. \mmax{This expression treats $R_\mathrm{thermal}$ and $C$ as constant parameters. In practice, the thermophysical properties of the wire and surrounding materials vary over the temperature range examined.}

For a TPP with $L=8$~mm, $D=6$~mm, electrical power of $P_\mathrm{el}=4.8$~W and a pulse duration of $t_p=75$~ms, the wire temperature reached a peak value of $T_\mathrm{wire}=1090~^\circ$C, corresponding to an inferred air temperature of $97~^\circ$C, a peak membrane force of $750$~mN, and a free displacement of $0.96$~mm. Following termination of the pulse, the system returned toward its initial state with a time constant of $\tau=110$~ms, obtained by fitting Eq.~\ref{eq:T_wire} to the cooling transient ($r^2=0.98$). Consistent with Eq.~\ref{eq:T_wire}, the peak force increased monotonically with both pulse duration and electrical power (Fig.~\ref{fig:mechanical}B-C).

To enable a comparison across geometries, we parameterized the power level by the electrical power dissipated per unit wire length, $\rho=P_\mathrm{el}/L_T$. Here, $L_T = 2L+1$ is the total wire length, accounting for routing across the cavity and back to the solder holes. In practice, the pair $(\rho, t_p)$ largely determines the peak wire temperature, and therefore provides a geometry-agnostic parameterization of the actuator's viable operating regime.

Using this parametrization, we examined how actuator geometry influences peak force output when driven at a power per unit length of $\rho = 0.5$~W/mm. Across TPPs with varying cavity length $L$ and membrane aperture diameter $D$, peak force increased monotonically with both parameters and reached a maximum measured value of $1.07$~N (Fig.~\ref{fig:mechanical}D). 

We next sought to identify the usable driving regime of the actuators. Owing to the low thermal mass of the NiCr wire, excessive drive conditions can rapidly result in wire melting and actuator failure. Specifically, we drove actuators with $L=4, 6,$ and $8$~mm to failure over a range of pulse durations $t_p$ and power per unit length values $\rho$ (Fig.~\ref{fig:mechanical}E). For each actuator, we identified the minimum pulse duration $t_p$ that caused wire failure, which occurred via melting near $T_\mathrm{wire}=1400~^\circ$C \cite{alloywire2016nicr}. Across lengths $L$ and driving levels $\rho$, the identified failure boundary clearly circumscribes the usable operating envelope of the TPPs (Fig.~\ref{fig:mechanical}E, shaded green area) -- a constraint surface that we parametrically fit to the observed data (see Appendix \ref{appx:fit} for details).


\subsection{Characterization During Cyclic Stimulation}

\begin{figure}
\begin{center}
\includegraphics[width = 85 mm]{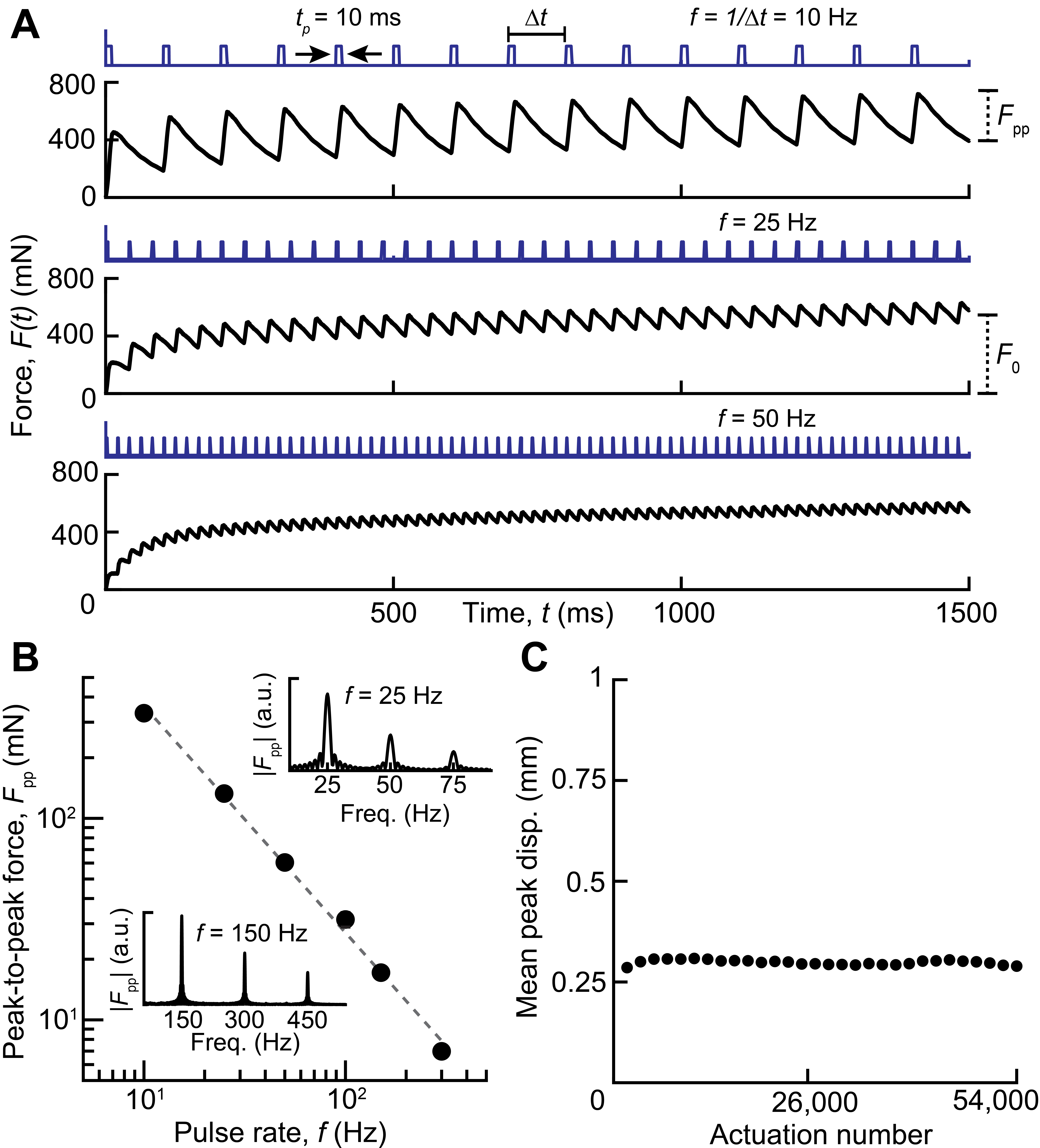}
\end{center}
\caption{Characterization of TPPs under cyclic operation. A) Isometric force responses to pulse trains at pulse rates $f = 10$, $25$, and $50$~Hz (top to bottom) at duty cycle $t_p/\Delta t=0.1$. B) Peak-to-peak force component $F_{pp}$ as a function of pulse rate $f$ with duty cycle $t_p/\Delta t=0.1$. Regression fit: $\log_{10}(F_{pp})=\alpha\log_{10}(f)+\beta$, $\alpha=-1.12$, $\beta=3.68$, $r^2>0.99$. Insets show the magnitude spectra of the measured force signals for pulse rates of 25 and 150 Hz.  (C) Mean free displacement is consistent over 54,000 actuations. Each dot corresponds to the mean of 90 peaks.
}
\label{fig:dynamic}
\end{figure}

During interactions with everyday objects, the human hand is exposed to tactile stimuli spanning a wide range of temporal frequencies. In particular, the skin is sensitive to transient displacements on timescales of approximately 1–10 ms \cite{morioka2005thresholds}, making the dynamic response of a tactile actuator a critical performance consideration. As shown in Fig.~\ref{fig:dynamic}, the TPPs support cyclic operation over a wide range of pulse rates, with pulse durations between 5 and 100 ms, despite a longer thermal relaxation time of $\tau = 110$ ms. Under such cyclic excitation, the force response can be decomposed into a slowly varying offset component, $F_0$, arising from residual heat accumulation in the cavity, and a pulse-synchronous component, $F_{pp}$, that tracks the applied pulse rate \cite{linnander2025tactile,linnander2026hled,mazzotta2025thermopneumatic}. This separation aligns well with tactile perception, which is known to be more sensitive to transient, high-frequency components of mechanical stimulation than to low-frequency or quasi-static inputs \cite{morioka2005thresholds}.

Given the heightened sensitivity of the tactile system to transient, high-frequency components of mechanical stimulation, the peak-to-peak force component, $F_{pp}$, is the primary metric of interest. We measured $F_{pp}$ for pulse rates $f$ between 10 and 300~Hz (Fig.~\ref{fig:dynamic}B), over which $F_{pp}$ ranged from 332~mN to 7.0~mN. Across this interval, the peak-to-peak force exhibited an approximately inverse dependence on frequency, with a slope near $-1$ in log–log space. This scaling is consistent with the fact that, for fixed electrical power $P_\mathrm{el}$ and duty cycle $t_p/\Delta t$ (where $\Delta t$ is the pulse period), the energy delivered per pulse decreases with increasing pulse rate, $E \propto 1/f$. Inset magnitude spectra (high-pass filtered) show strong spectral peaks at the drive frequency and its harmonics, consistent with the sawtooth-like temporal force waveform (insets in Fig.~\ref{fig:dynamic}B).

We evaluated the reliability of the actuator over 54,000 actuation cycles and observed no meaningful degradation in dynamic performance (Fig.~\ref{fig:dynamic}C). The driving parameters used, $\rho=0.42$~W/mm and $t_p=19$~ms, correspond to an inferred peak wire temperature of approximately $800~^\circ$C and air temperature of $80~^\circ$C, reached during each pulse. At this temperature, NiCr exhibits strong resistance to oxidation, supporting stable operation under extended cyclic actuation \cite{zhou2003nicr}.

\subsection{Surface Temperature Characterization}

\begin{figure}[b]
\begin{center}
\includegraphics[width = 85 mm]{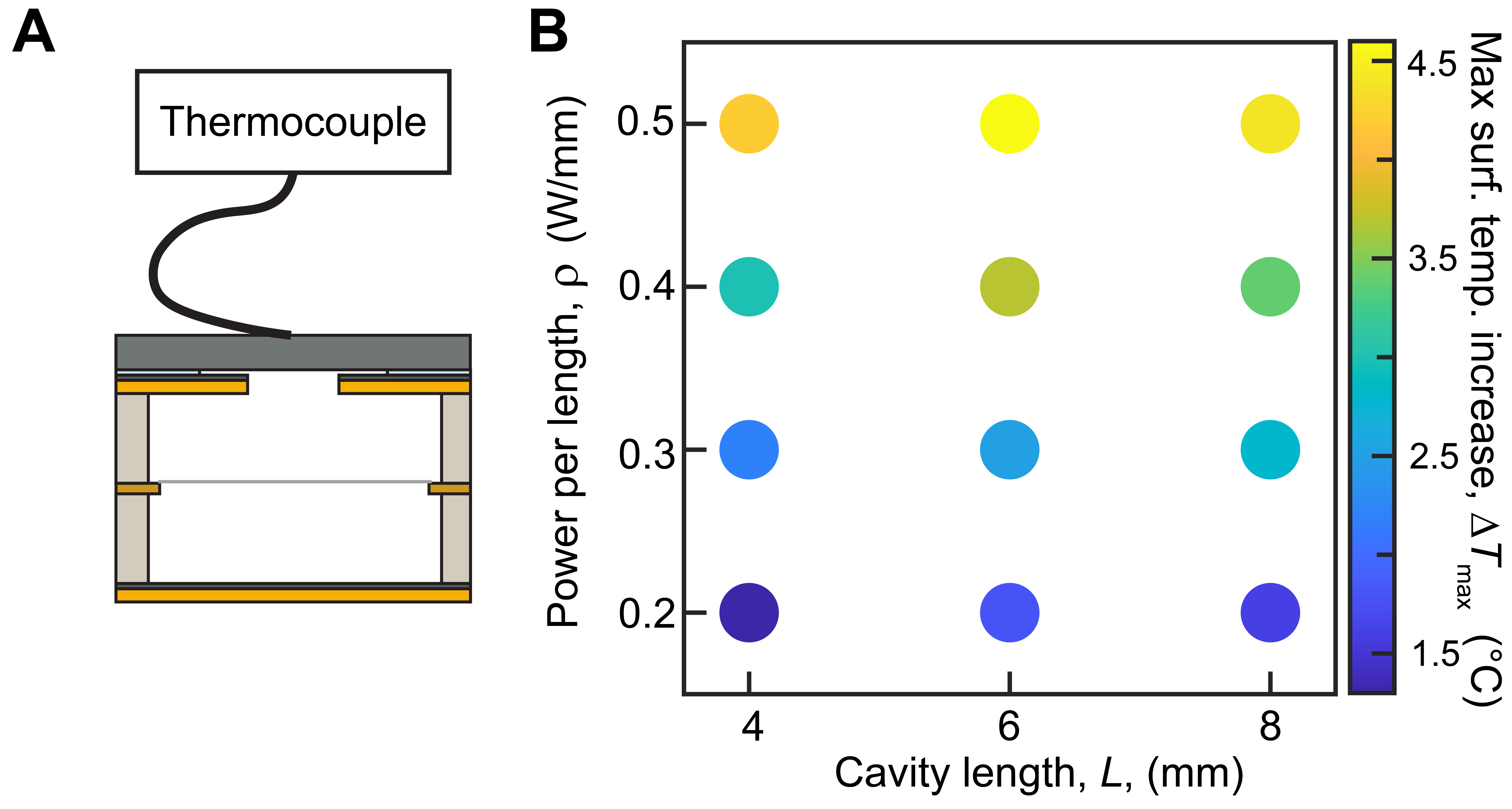}
\end{center}
\caption{A) Surface temperature measurements of TPPs. (b) Max temperatures measured during a 5-s actuation window for varying $L$ and $\rho$ at a duty cycle of $t_p/\Delta t=0.2$.}
\label{fig:thermal}
\end{figure}

We measured the maximum surface temperature increase, $\Delta T_\mathrm{surf}$, at the center of the membrane, and found that surface heating remained low for the conditions examined (Fig.~\ref{fig:thermal}). Across all conditions tested, the maximum membrane temperature rise was $\Delta T_\mathrm{surf}=4.6$~$^\circ$C, comparable to the temperature increase induced by a finger in contact with the surface \cite{linnander2025tactile}. Measurements were performed during continuous pulsed operation over a 5-s window with duty cycle $t_p/\Delta t=0.2$ and pulse rate $f=20$~Hz, while varying cavity length $L$ and power per unit length $\rho$. With duty cycle held constant, $\Delta T_\mathrm{surf}$ varied little across geometries and was mainly influenced by $\rho$. To place these measurements in context, tactile actuators are rarely driven continuously at high power for more than brief intervals, as sustained stimulation can be uncomfortable and subject to sensory adaptation \cite{verrillo1977effect}. Consistent with this, modern consumer devices typically employ short vibrational bursts rather than prolonged actuation.

\section{Perceptual Experiments}

We conducted two perceptual experiments to evaluate tactile feedback delivered by the thermopneumatic pixels (TPPs).   The first study characterized perceived intensity as a function of power, $P_\mathrm{el}$. The second study consisted of a localization task where the user identified which TPP on their finger pad was active. 

We recruited ten people who volunteered to participate without compensation (4 female; finger pad width 1.3 to 1.9 cm; ages 20 to 29). All participants provided their written informed consent, and the experimental protocol was approved by the human subjects review board at the authors’ institution
. At the start of each experiment, before data collection, participants experienced several stimuli and practiced entering responses for familiarization. No feedback about responses was supplied to participants during this phase, and the data were not used for analysis. 

\begin{figure}
\begin{center}
\includegraphics[width = 85 mm]{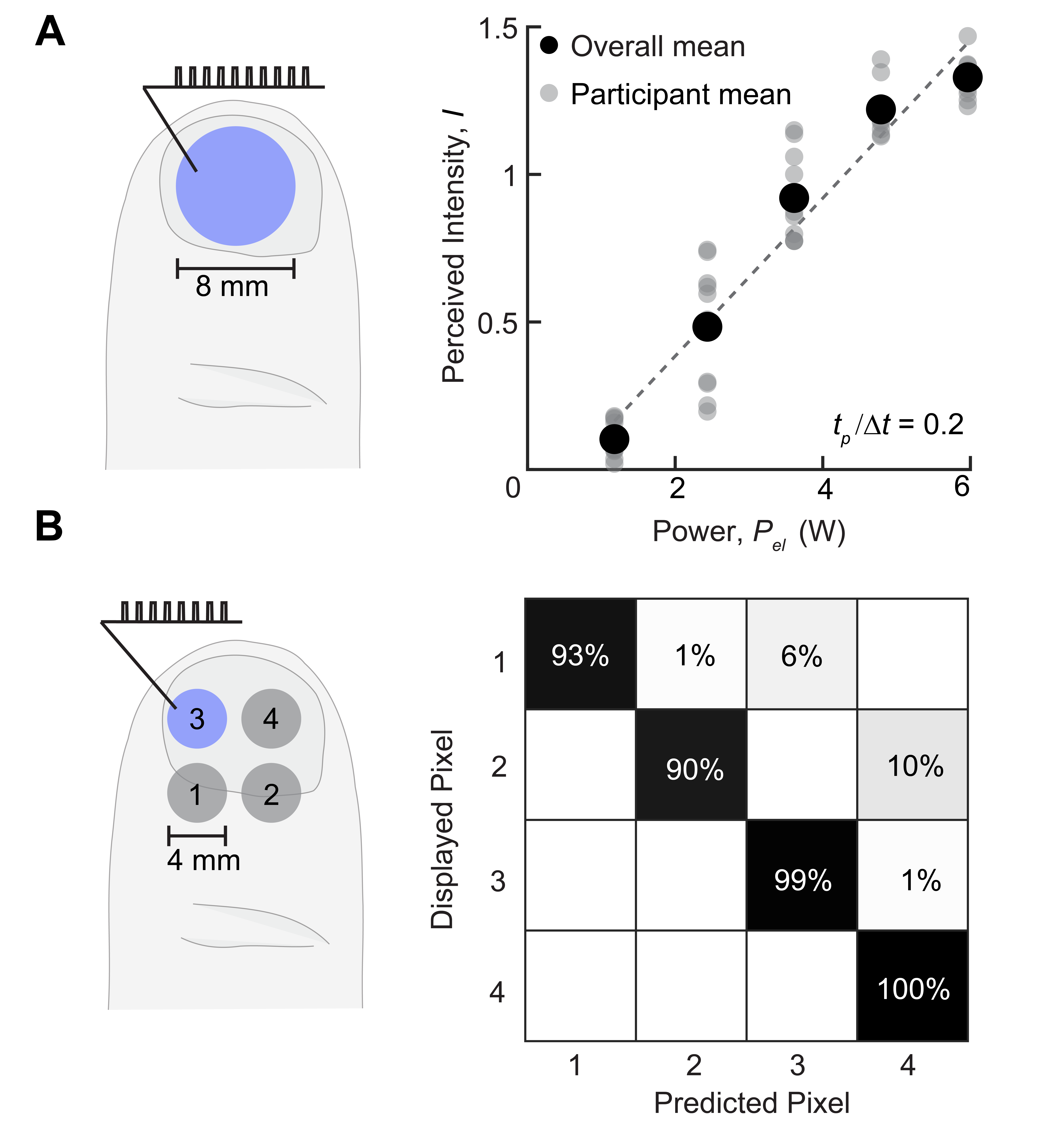}
\end{center}
\caption{Perceptual Evaluation. A) Experiment 1: Perceived intensity as a function of power, measured using the method of magnitude scaling. Regression fit: $I = \alpha P_\mathrm{el}+\beta$, $\alpha = 0.2677$, $\beta = -0.151$, $r^2 = 0.97$. B) Experiment 2: Spatial localization of active TPP on the finger pad.}
\label{fig:perceptual}
\end{figure}

In the first experiment, we assessed perceived tactile intensity $I$ as a function of dissipated electrical power $P_\mathrm{el}$, finding that perceived intensity, $I(P_\mathrm{el})$, increased linearly with power (Fig.~\ref{fig:perceptual}A). The experiment used the psychophysical method of magnitude estimation \cite{jones2013psychophysics,stevens1958problems}. This experiment used the TPP module with a single pixel ($L = 10$~mm and $D = 8$~mm; Fig. 1G).  The stimuli were pulse trains ($f~=~20$~Hz, $t_p/\Delta t~=~0.2$, duration = 0.5 s) at each of 5 different power levels ($P_\mathrm{el} = $ 1.2 to 6~W).  Each stimulus was presented 7 times.  Participants rated the intensity of each stimulus on a continuous numerical scale on a computer interface. The extremes of the scale were labeled ``no sensation'' and ``strongest imaginable''. Perceived intensity, $I(P_\mathrm{el})$, was computed for each power level $P_\mathrm{el}$  by averaging the normalized geometric mean of responses across all participants \cite{jones2013psychophysics}. Using these results, a desired intensity level can be generated by driving the TPP with the corresponding power.  
In experiment 2, participants reported which single TPP was actuated out of four TPPs ($L = 4$~mm, $D = 4$~mm) arranged in a grid under their finger pad (Fig.~\ref{fig:perceptual}B). They responded correctly on
95.5\% of trials. The stimuli were pulse trains ($f~=~20$~Hz, $t_p/\Delta t~=~0.2$, duration = 0.5 s, $P_\mathrm{el} = 2.8$~W) with 28 trials per participant (4 locations x 7 repetitions). In post-experiment surveys, nine participants reported feeling no warmth from the TPP modules, while one participant reported feeling slight warmth.  

\section{Conclusion}


In this work, we presented thermopneumatic pixels (TPPs), a compact tactile actuation platform for wearable and surface-based haptic systems that delivers computationally controllable, localized tactile output in thin, planar assemblies with no internal moving parts that can be driven at low-voltages. Using a layered construction based on inexpensive, commodity materials, TPPs facilitate facile fabrication and integration.  The forces and displacements they produce can be adjusted over an order of magnitude or more by geometric scaling of the assembly.  We demonstrated devices that generate forces exceeding 1 N and displacements on the millimeter-scale, at response times on the order of 5 to 100 ms.  They also admit cyclic, oscillating actuation modes across a broad operating range of frequencies. We demonstrated concurrent operation of multiple modules through a simple, scalable driver architecture and showed through perceptual experiments that the resulting tactile signals are salient and spatially localizable.  Our results also characterize the scaling of perceived intensity with driving power. Together, these results establish TPPs as a practical and adaptable approach for embedding localized tactile feedback into compact interactive devices and other application-facing haptic interfaces.

\appendices

\section{Hardware Components for Tactile Display} \label{appx:tables}

\begin{table}[H]
\centering
\caption{Electrical components used in the custom driving circuit for controlling up to 40 TPPs.}
\label{tab:electricalComponents}
\begin{tabular}{|c|c|c|c|}
\hline
\textbf{Component} & \textbf{Part Number} & \textbf{Quantity} & \textbf{Cost}  \\
\hline
Microcontroller & Teensy 4.1 & 1 & \$33.6 \\
MOSFET & AO3400A & 40 & \$1.97 \\
Resistor & 0402WGF3302TCE & 40 & \$0.03 \\
Connector & 691137710002 & 1 & \$0.7 \\
Connector & 522070760 & 10 & \$2.79 \\
Connector & PPPC241LFBN-RC & 2 & \$4.4 \\
\hline
\multicolumn{3}{|r|}{\textbf{\rule{0pt}{2.2ex}Total Cost}} & \textbf{\rule{0pt}{2.2ex}\$43.49} \\
\hline
\end{tabular}
\end{table}

\begin{table}[H]
\centering
\caption{Components used for one TPP module}
\label{tab:electricalComponents}
\begin{tabular}{|c|c|c|}
\hline
\textbf{Component} & \textbf{Part Number}  & \textbf{Unit Cost}  \\
\hline
FPC & JLCPCB Custom Design &  \$2.2 \\
NiCr Wire & Wire Optim 354-23 &  \$0.01 \\
Polyimide & McMaster-Carr 2271K73 & \$0.24 \\
Polysiloxane & McMaster-Carr 86915K24 & \$0.24 \\
Adhesive & 3M 2477 &  \$0.01 \\
PDMS & McMaster-Carr 86435K41 & \$0.12 \\
PGS & HPMS Graphite HGS-012-12A & \$0.2 \\
\hline
\multicolumn{2}{|r|}{\textbf{\rule{0pt}{2.2ex}Total Cost}} & \textbf{\rule{0pt}{2.2ex}\$3.02} \\
\hline
\end{tabular}
\end{table}

\section{Wire Temperature from Resistance Measurement} \label{appx:wire_temp}

The wire temperature, $T_\mathrm{wire}(t)$, in Fig.~\ref{fig:mechanical}A was computed from resistance measurements of the NiCr wire during actuation. A shunt resistor ($R_\mathrm{shunt}=0.22~\Omega$) was placed in series with the TPP, and the voltage drop across it, $V_\mathrm{shunt}(t)$, was measured during actuation. Assuming that the driving voltage $V_+$ was constant, and that there was a negligible temperature rise in the shunt and remaining circuitry ($R_\mathrm{circuit}=2.1~\Omega$), we obtained the total series resistance via the inverse voltage divider relation,
\[
R_\mathrm{tot}(t) = \frac{V_+ R_\mathrm{shunt}}{V_\mathrm{shunt}(t)}.
\]
The instantaneous wire resistance was then calculated as $R_\mathrm{wire}(t)=R_\mathrm{tot}(t)-R_\mathrm{shunt}-R_\mathrm{circuit}$. The relative resistivity change was referenced to the initial wire resistance, and the corresponding temperature rise was estimated by linearly interpolating tabulated resistivity–temperature data for NiCr \cite{alloywire2016nicr}. The resulting temperature trace was smoothed using a moving-average filter and downsampled for visualization.

\section{Operating Envelope Fit} \label{appx:fit}

The experimentally observed operating envelope boundary in Fig.~\ref{fig:mechanical}E was fit using the lumped-parameter thermal model for the NiCr heating element (Eq.~\ref{eq:T_wire}). Solving Eq.~\ref{eq:T_wire} for $P_\mathrm{el}$ and normalizing by the total wire length $L_T$ yields the power per unit length, $\rho = P_\mathrm{el}/L_T$, in terms of pulse duration $t_p$,
\begin{equation}
\rho(t_p) = \frac{T_\mathrm{fail}}{a\!\left(1 - e^{-t_p/b}\right)}
\label{eq:rho_fit}
\end{equation}
Here, the wire failure temperature, $T_\mathrm{fail} = 1400~^\circ$C, was set to the melting temperature of NiCr. The fitting parameters $a$ and $b$ capture the length-scaled thermal properties of the heating element, with $a = R_\mathrm{thermal} L_T$ representing the thermal resistance and $b = C/L_T$ the thermal mass. The analytic fit closely aligned with the measured failure points ($a = 6601$~mm K/W, $b = 6.51~\mu$J/mm K, $r^2 = 0.99$), delineating a region of stable operation (green) from drive conditions that lead to failure (red) (Fig. \ref{fig:mechanical}E).


\bibliographystyle{IEEEtran}
\bibliography{IEEEabrv,refs}

@article{stevens1958problems,
  title={Problems and methods of psychophysics.},
  author={Stevens, Stanley Smith},
  journal={Psychological bulletin},
  volume={55},
  number={4},
  pages={177},
  year={1958},
  publisher={American Psychological Association}
}

@article{copeland2010identification,
  title={Identification of the optimum resolution specification for a haptic graphic display},
  author={Copeland, Damian and Finlay, Janet},
  journal={Interacting with Computers},
  volume={22},
  number={2},
  pages={98--106},
  year={2010},
  publisher={Oxford University Press Oxford, UK}
}

@book{kern2023engineering,
  title={Engineering haptic devices},
  author={Kern, Thorsten A and Hatzfeld, Christian and Abbasimoshaei, Alireza},
  year={2023},
  publisher={Springer Nature}
}

@article{craig1987vibrotactile,
  title={Vibrotactile masking and the persistence of tactual features},
  author={Craig, James C and Evans, Paul M},
  journal={Perception and psychophysics},
  volume={42},
  number={4},
  pages={309--317},
  year={1987},
  publisher={Springer}
}

@book{jones2006human,
  title={Human hand function},
  author={Jones, Lynette and Lederman, Susan},
  year={2006},
  publisher={Oxford University Press}
}

@article{choi2012vibrotactile,
  title={Vibrotactile display: Perception, technology, and applications},
  author={Choi, Seungmoon and Kuchenbecker, Katherine J},
  journal={Proceedings of the IEEE},
  volume={101},
  number={9},
  pages={2093--2104},
  year={2012},
  publisher={IEEE}
}

@inproceedings{goetz2023dynamic,
  title={Dynamic Feedback in Wave-Mediated Surface Haptics: A Modular Platform},
  author={Goetz, Dustin and Reardon, Gregory and Linnander, Max and Visell, Yon},
  booktitle={2023 IEEE World Haptics Conference (WHC)},
  pages={107--113},
  year={2023},
  organization={IEEE}
}

@inproceedings{shen2023fluidreality,
author = {Shen, Vivian and Rae-Grant, Tucker and Mullenbach, Joe and Harrison, Chris and Shultz, Craig},
title = {Fluid Reality: High-Resolution, Untethered Haptic Gloves using Electroosmotic Pump Arrays},
year = {2023},
isbn = {9798400701320},
publisher = {Association for Computing Machinery},
address = {New York, NY, USA},
doi = {10.1145/3586183.3606771},
booktitle = {Proceedings of the 36th Annual ACM Symposium on User Interface Software and Technology},
articleno = {8},
numpages = {20},
location = {, San Francisco, CA, USA, },
series = {UIST '23}
}

@inproceedings{shultz2023flatpanel,
author = {Shultz, Craig and Harrison, Chris},
title = {Flat Panel Haptics: Embedded Electroosmotic Pumps for Scalable Shape Displays},
year = {2023},
isbn = {9781450394215},
publisher = {Association for Computing Machinery},
address = {New York, NY, USA},
doi = {10.1145/3544548.3581547},
booktitle = {Proceedings of the 2023 CHI Conference on Human Factors in Computing Systems},
articleno = {745},
numpages = {16},
location = {Hamburg, Germany, },
series = {CHI '23}
}

@article{leroy2020haxel,
author = {Leroy, Edouard and Hinchet, Ronan and Shea, Herbert},
title = {Multimode Hydraulically Amplified Electrostatic Actuators for Wearable Haptics},
journal = {Advanced Materials},
volume = {32},
number = {36},
pages = {2002564},
doi = {https://doi.org/10.1002/adma.202002564},
year = {2020}
}

@article{grasso2023haxel,
author = {Grasso, Giulio and Rosset, Samuel and Shea, Herbert},
title = {Fully 3D-Printed, Stretchable, and Conformable Haptic Interfaces},
journal = {Advanced Functional Materials},
volume = {33},
number = {20},
pages = {2213821},
doi = {https://doi.org/10.1002/adfm.202213821},
year = {2023}
}

@article{leroy2023haxel,
author = {Leroy, Edouard and Shea, Herbert},
title = {Hydraulically Amplified Electrostatic Taxels (HAXELs) for Full Body Haptics},
journal = {Advanced Materials Technologies},
volume = {8},
number = {16},
pages = {2300242},
doi = {https://doi.org/10.1002/admt.202300242},
year = {2023}
}

@article{hwang2021bimorph,
author = {Hwang, Inwook and Kim, Hyeong Jun and Mun, Seongcheol and Yun, Sungryul and Kang, Tae June},
title = {A Light-Driven Vibrotactile Actuator with a Polymer Bimorph Film for Localized Haptic Rendering},
journal = {ACS Applied Materials and Interfaces},
volume = {13},
number = {5},
pages = {6597-6605},
year = {2021},
doi = {10.1021/acsami.0c19003},
}

@article{torras2014optomechanical,
title = {Tactile device based on opto-mechanical actuation of liquid crystal elastomers},
journal = {Sensors and Actuators A: Physical},
volume = {208},
pages = {104-112},
year = {2014},
issn = {0924-4247},
doi = {https://doi.org/10.1016/j.sna.2014.01.012},
author = {N. Torras and K.E. Zinoviev and C.J. Camargo and Eva M. Campo and H. Campanella and J. Esteve and J.E. Marshall and E.M. Terentjev and M. Omastová and I. Krupa and P. Teplický and B. Mamojka and P. Bruns and B. Roeder and M. Vallribera and R. Malet and S. Zuffanelli and V. Soler and J. Roig and N. Walker and D. Wenn and F. Vossen and F.M.H. Crompvoets},
}

@article{camargo2012opticalbraille,
doi = {10.1088/0960-1317/22/7/075009},
url = {https://dx.doi.org/10.1088/0960-1317/22/7/075009},
year = {2012},
month = {jun},
publisher = {IOP Publishing},
volume = {22},
number = {7},
pages = {075009},
author = {C J Camargo and H Campanella and J E Marshall and N Torras and K Zinoviev and E M Terentjev and J Esteve},
title = {Batch fabrication of optical actuators using nanotube–elastomer composites towards refreshable Braille displays},
journal = {Journal of Micromechanics and Microengineering},
}

@article{kastor2023ferrofluidic,
title = {Ferrofluid electromagnetic actuators for high-fidelity haptic feedback},
journal = {Sensors and Actuators A: Physical},
volume = {355},
pages = {114252},
year = {2023},
issn = {0924-4247},
doi = {https://doi.org/10.1016/j.sna.2023.114252},
author = {Nikolas Kastor and Bharat Dandu and Vedad Bassari and Gregory Reardon and Yon Visell},
}

@article{streque2012emactuator,
doi = {10.1088/0960-1317/22/9/095020},
year = {2012},
month = {aug},
publisher = {IOP Publishing},
volume = {22},
number = {9},
pages = {095020},
author = {J Streque and A Talbi and P Pernod and V Preobrazhensky},
title = {Pulse-driven magnetostatic micro-actuator array based on ultrasoft elastomeric membranes for active surface applications},
journal = {Journal of Micromechanics and Microengineering},
}

@article{besse2017fea,
author = {Besse, Nadine and Rosset, Samuel and Zarate, Juan Jose and Shea, Herbert},
title = {Flexible Active Skin: Large Reconfigurable Arrays of Individually Addressed Shape Memory Polymer Actuators},
journal = {Advanced Materials Technologies},
year = {2017},
volume = {2},
number = {10},
pages = {1700102},
doi = {https://doi.org/10.1002/admt.201700102}
}

@ARTICLE{wu2012pneumatic,
author={Wu, Xiaosong and Kim, Seong-Hyok and Zhu, Haihong and Ji, Chang-Hyeon and Allen, Mark G.},
journal={Journal of Microelectromechanical Systems}, 
title={A Refreshable Braille Cell Based on Pneumatic Microbubble Actuators}, 
year={2012},
volume={21},
number={4},
pages={908-916},
doi={10.1109/JMEMS.2012.2190043}}

@ARTICLE{jones2013psychophysics,
  author={Jones, Lynette A. and Tan, Hong Z.},
  journal={IEEE Transactions on Haptics}, 
  title={Application of Psychophysical Techniques to Haptic Research}, 
  year={2013},
  volume={6},
  number={3},
  pages={268-284},
  doi={10.1109/TOH.2012.74}}

@article{biswas2019materials,
author = {Biswas, Shantonu and Visell, Yon},
title = {Emerging Material Technologies for Haptics},
journal = {Advanced Materials Technologies},
volume = {4},
number = {4},
pages = {1900042},
doi = {https://doi.org/10.1002/admt.201900042},
year = {2019}
}

@ARTICLE{hiraki2020laserpouch,
  author={Hiraki, Takefumi and Nakahara, Kenichi and Narumi, Koya and Niiyama, Ryuma and Kida, Noriaki and Takamura, Naoki and Okamoto, Hiroshi and Kawahara, Yoshihiro},
  journal={IEEE Robotics and Automation Letters}, 
  title={Laser Pouch Motors: Selective and Wireless Activation of Soft Actuators by Laser-Powered Liquid-to-Gas Phase Change}, 
  year={2020},
  volume={5},
  number={3},
  pages={4180-4187},
  doi={10.1109/LRA.2020.2982864}}

@INPROCEEDINGS{russomanno2015pneumatic,
  author={Russomanno, Alexander and Gillespie, R. Brent and O'Modhrain, Sile and Burns, Mark},
  booktitle={2015 IEEE World Haptics Conference (WHC)}, 
  title={The design of pressure-controlled valves for a refreshable tactile display}, 
  year={2015},
  volume={},
  number={},
  pages={177-182},
  doi={10.1109/WHC.2015.7177710}}

@INPROCEEDINGS{russomanno2017pneumatic,
  author={Russomanno, Alex and Xu, Zhentao and O'Modhrain, Sile and Gillespie, Brent},
  booktitle={2017 IEEE World Haptics Conference (WHC)}, 
  title={A pneu shape display: Physical buttons with programmable touch response}, 
  year={2017},
  volume={},
  number={},
  pages={641-646},
  doi={10.1109/WHC.2017.7989976}}

@inproceedings{lining2013pneumatic,
author = {Yao, Lining and Niiyama, Ryuma and Ou, Jifei and Follmer, Sean and Della Silva, Clark and Ishii, Hiroshi},
title = {PneUI: pneumatically actuated soft composite materials for shape changing interfaces},
year = {2013},
isbn = {9781450322683},
publisher = {Association for Computing Machinery},
address = {New York, NY, USA},
url = {https://doi.org/10.1145/2501988.2502037},
doi = {10.1145/2501988.2502037},
booktitle = {Proceedings of the 26th Annual ACM Symposium on User Interface Software and Technology},
pages = {13–22},
numpages = {10},
location = {St. Andrews, Scotland, United Kingdom},
series = {UIST '13}
}

@article{wang2019multimodal,
  title={Multimodal haptic display for virtual reality: A survey},
  author={Wang, Dangxiao and Ohnishi, Kouhei and Xu, Weiliang},
  journal={IEEE Transactions on Industrial Electronics},
  volume={67},
  number={1},
  pages={610--623},
  year={2019},
  publisher={IEEE}
}

@techreport{colgate2013haptic,
  title={Haptic Interface for Vehicular Touch Screens},
  author={Colgate, J Edward and Peshkin, Michael A and others},
  year={2013},
  publisher={US Department of Transportation},
  institution={Center for the Commercialization of Innovative Transportation Technologies}
}

@article{qiu2018polymerdisplay,
author = {Qiu, Yu and Lu, Zhiyun and Pei, Qibing},
title = {Refreshable Tactile Display Based on a Bistable Electroactive Polymer and a Stretchable Serpentine Joule Heating Electrode},
journal = {ACS Applied Materials and Interfaces},
volume = {10},
number = {29},
pages = {24807-24815},
year = {2018},
doi = {10.1021/acsami.8b07020}
}

@article{zhao2018dea,
author = {Zhao, Huichan and Hussain, Aftab M. and Duduta, Mihai and Vogt, Daniel M. and Wood, Robert J. and Clarke, David R.},
title = {Compact Dielectric Elastomer Linear Actuators},
journal = {Advanced Functional Materials},
volume = {28},
number = {42},
pages = {1804328},
doi = {https://doi.org/10.1002/adfm.201804328},
url = {https://onlinelibrary.wiley.com/doi/abs/10.1002/adfm.201804328},
eprint = {https://onlinelibrary.wiley.com/doi/pdf/10.1002/adfm.201804328},
year = {2018}
}

@article{ji2021dea,
author = {Ji, Xiaobin and Liu, Xinchang and Cacucciolo, Vito and Civet, Yoan and El Haitami, Alae and Cantin, Sophie and Perriard, Yves and Shea, Herbert},
title = {Untethered Feel-Through Haptics Using 18-µm Thick Dielectric Elastomer Actuators},
journal = {Advanced Functional Materials},
volume = {31},
number = {39},
pages = {2006639},
doi = {https://doi.org/10.1002/adfm.202006639},
url = {https://onlinelibrary.wiley.com/doi/abs/10.1002/adfm.202006639},
eprint = {https://onlinelibrary.wiley.com/doi/pdf/10.1002/adfm.202006639},
year = {2021}
}

@inproceedings{strasnick2016magneticactuators,
author = {Strasnick, Evan and Follmer, Sean},
title = {Applications of Switchable Permanent Magnetic Actuators in Shape Change and Tactile Display},
year = {2016},
isbn = {9781450345316},
publisher = {Association for Computing Machinery},
address = {New York, NY, USA},
url = {https://doi.org/10.1145/2984751.2985728},
doi = {10.1145/2984751.2985728},
booktitle = {Adjunct Proceedings of the 29th Annual ACM Symposium on User Interface Software and Technology},
pages = {123–125},
numpages = {3},
location = {Tokyo, Japan},
series = {UIST '16 Adjunct}
}

@ARTICLE{kato2007polymeractuator,
  author={Kato, Yusaku and Sekitani, Tsuyoshi and Takamiya, Makoto and Doi, Masao and Asaka, Kinji and Sakurai, Takayasu and Someya, Takao},
  journal={IEEE Transactions on Electron Devices}, 
  title={Sheet-Type Braille Displays by Integrating Organic Field-Effect Transistors and Polymeric Actuators}, 
  year={2007},
  volume={54},
  number={2},
  pages={202-209},
  doi={10.1109/TED.2006.888678}}

@ARTICLE{koo2008soft,
  author={Koo, Ig Mo and Jung, Kwangmok and Koo, Ja Choon and Nam, Jae-Do and Lee, Young Kwan and Choi, Hyouk Ryeol},
  journal={IEEE Transactions on Robotics}, 
  title={Development of Soft-Actuator-Based Wearable Tactile Display}, 
  year={2008},
  volume={24},
  number={3},
  pages={549-558},
  doi={10.1109/TRO.2008.921561}}

@ARTICLE{uramune2022hapouch,
  author={Uramune, Ryusei and Ishizuka, Hiroki and Hiraki, Takefumi and Kawahara, Yoshihiro and Ikeda, Sei and Oshiro, Osamu},
  journal={IEEE Access}, 
  title={HaPouch: A Miniaturized, Soft, and Wearable Haptic Display Device Using a Liquid-to-Gas Phase Change Actuator}, 
  year={2022},
  volume={10},
  number={},
  pages={16830-16842},
  doi={10.1109/ACCESS.2022.3141385}}

@ARTICLE{ueno2020phasechange,
  author={Ueno, Soichiro and Monnai, Yasuaki},
  journal={IEEE Robotics and Automation Letters}, 
  title={Wireless Soft Actuator Based on Liquid-Gas Phase Transition Controlled by Millimeter-Wave Irradiation}, 
  year={2020},
  volume={5},
  number={4},
  pages={6483-6488},
  doi={10.1109/LRA.2020.3013847}}

@article{verrillo1977effect,
  title={Effect of prior stimulation on vibrotactile thresholds},
  author={Verrillo, Ronald T and Gescheider, George A},
  journal={Sensory Processes},
  volume={1},
  number={4},
  pages={292--300},
  year={1977},
  publisher={Springer}
}

@article{morioka2005thresholds,
  title={Thresholds for the perception of hand-transmitted vibration: dependence on contact area and contact location},
  author={Morioka, Miyuki and Griffin, Michael J},
  journal={Somatosensory and Motor Research},
  volume={22},
  number={4},
  pages={281--297},
  year={2005},
  publisher={Taylor and Francis},
  doi={10.1080/08990220500420400},
  pmid={16503581}
}

@article{linnander2025tactile,
author = {Max Linnander  and Dustin Goetz  and Gregory Reardon  and Vijay Kumar  and Elliot Hawkes  and Yon Visell },
title = {Tactile displays driven by projected light},
journal = {Science Robotics},
volume = {10},
number = {107},
pages = {eadv1383},
year = {2025},
doi = {10.1126/scirobotics.adv1383},
URL = {https://www.science.org/doi/abs/10.1126/scirobotics.adv1383},
eprint = {https://www.science.org/doi/pdf/10.1126/scirobotics.adv1383},
}

@article{mazzotta2025thermopneumatic,
  title={Low-voltage wearable tactile display with thermo-pneumatic actuation},
  author={Mazzotta, A. and Taccola, S. and Cesini, I. and others},
  journal={npj Flexible Electronics},
  volume={9},
  number={70},
  year={2025},
  publisher={Nature Publishing Group},
  doi={10.1038/s41528-025-00426-3},
  url={https://doi.org/10.1038/s41528-025-00426-3}
}

@article{mazzotta2023thermopneumatic,
author = {Mazzotta, Arianna and Mattoli, Virgilio},
title = {Ultrathin Conformable Electronic Tattoo for Tactile Sensations},
journal = {Advanced Electronic Materials},
volume = {9},
number = {9},
pages = {2201327},
doi = {https://doi.org/10.1002/aelm.202201327},
year = {2023}
}

@article{mitchell2022HV,
author = {Mitchell, Shane K. and Martin, Trent and Keplinger, Christoph},
title = {A Pocket-Sized Ten-Channel High Voltage Power Supply for Soft Electrostatic Actuators},
journal = {Advanced Materials Technologies},
volume = {7},
number = {8},
pages = {2101469},
doi = {https://doi.org/10.1002/admt.202101469},
url = {https://advanced.onlinelibrary.wiley.com/doi/abs/10.1002/admt.202101469},
year = {2022}
}

@article{linnander2026hled,
	author = {Linnander, Max and Visell, Yon},
	title = {Haptic light-emitting diodes: Miniature, luminous tactile actuators},
	journal = {Applied Physics Letters},
	volume = {128},
	number = {15},
	pages = {153505},
	year = {2026},
	month = {04},
	issn = {0003-6951},
	doi = {10.1063/5.0324354},
	url = {https://doi.org/10.1063/5.0324354},
	eprint = {https://pubs.aip.org/aip/apl/article-pdf/doi/10.1063/5.0324354/20978270/153505_1_5.0324354.pdf}
}

@article{zhou2003nicr,
    author = {Zhou, J. and Ohno, T. R. and Wolden, C. A.},
    title = {High-temperature stability of nichrome in reactive environments},
    journal = {Journal of Vacuum Science and Technology A},
    volume = {21},
    number = {3},
    pages = {756-761},
    year = {2003},
    month = {04},
    issn = {0734-2101},
    doi = {10.1116/1.1570834},
    url = {https://doi.org/10.1116/1.1570834},
    eprint = {https://pubs.aip.org/avs/jva/article-pdf/21/3/756/11024654/756_1_online.pdf},
}

@INPROCEEDINGS{tummala2024skinsource,
  author={Tummala, Neeli and Reardon, Gregory and Fani, Simone and Goetz, Dustin and Bianchi, Matteo and Visell, Yon},
  booktitle={2024 IEEE Haptics Symposium (HAPTICS)}, 
  title={SkinSource: A Data-Driven Toolbox for Predicting Touch-Elicited Vibrations in the Upper Limb}, 
  year={2024},
  volume={},
  number={},
  pages={53-60},
  doi={10.1109/HAPTICS59260.2024.10520852}
}

@inproceedings{huang2025vibraforge,
  author = {Huang, Bingjian and Ren, Siyi and Luo, Yuewen and Cheng, Qilong and Cai, Hanfeng and Sang, Yeqi and Sousa, Mauricio and Dietz, Paul H and Wigdor, Daniel},
  title = {VibraForge: A Scalable Prototyping Toolkit For Creating Spatialized Vibrotactile Feedback Systems},
  year = {2025},
  publisher = {Association for Computing Machinery},
  address = {New York, NY, USA},
  doi = {10.1145/3706598.3714273},
  booktitle = {Proceedings of the 2025 CHI Conference on Human Factors in Computing Systems},
  articleno = {1137},
  numpages = {18},
  series = {CHI '25}
}

@article{rokhmanova2025platform,
  title = {Open-Source Hardware and Software Platform for Vibrotactile Motion Guidance},
  journal = {Device},
  pages = {100966},
  year = {2025},
  month = oct,
  author = {Rokhmanova, Nataliya and Martus, Julian and Faulkner, Robert and Fiene, Jonathan and Kuchenbecker, Katherine J.},
  doi = {10.1016/j.device.2025.100966}
}

@techreport{alloywire2016nicr,
  author       = {{Alloy Wire International Ltd.}},
  title        = {80/20 Ni Cr Resistance Wire: Technical Datasheet AWS 180 Rev.1},
  institution  = {Alloy Wire International Ltd.},
  year         = {2016},
  address      = {Brierley Hill, West Midlands, UK},
  type         = {Technical Datasheet},
  number       = {AWS 180 Rev.1},
  note         = {Available at: https://www.alloywire.com}
}

@article{okamura2009medical,
  author    = {Allison M. Okamura},
  title     = {Haptic Feedback in Robot-Assisted Minimally Invasive Surgery},
  journal   = {Current Opinion in Urology},
  year      = {2009},
  volume     = {19},
  number     = {1},
  pages      = {102--107},
  month      = jan,
  doi        = {10.1097/MOU.0b013e32831a478c},
  pmid       = {19057225},
  pmcid      = {PMC2701448},
  issn       = {0963-0643},
  publisher  = {Lippincott Williams \& Wilkins}
}

@inproceedings{lee2004stylus,
author = {Lee, Johnny C. and Dietz, Paul H. and Leigh, Darren and Yerazunis, William S. and Hudson, Scott E.},
title = {Haptic pen: a tactile feedback stylus for touch screens},
year = {2004},
isbn = {1581139578},
publisher = {Association for Computing Machinery},
address = {New York, NY, USA},
url = {https://doi.org/10.1145/1029632.1029682},
doi = {10.1145/1029632.1029682},
booktitle = {Proceedings of the 17th Annual ACM Symposium on User Interface Software and Technology},
pages = {291–294},
numpages = {4},
location = {Santa Fe, NM, USA},
series = {UIST '04}
}

@ARTICLE{shan2024hydraulic,
	author={Shan, Boxue and Liu, Congying and Guo, Yuan and Wang, Yiheng and Guo, Weidong and Zhang, Yuru and Wang, Dangxiao},
	journal={IEEE Transactions on Haptics}, 
	title={A Multi-Layer Stacked Microfluidic Tactile Display With High Spatial Resolution}, 
	year={2024},
	volume={17},
	number={4},
	pages={546-556},
	doi={10.1109/TOH.2024.3367708}
}

@article{frediani2021eap,
	author = {Frediani, Gabriele and Boys, Hugh and Ghilardi, Michele and Poslad, Stefan and Busfield, James J. C. and Carpi, Federico},
	title = {A Soft Touch: Wearable Tactile Display of Softness Made of Electroactive Elastomers},
	journal = {Advanced Materials Technologies},
	volume = {6},
	number = {6},
	pages = {2100016},
	doi = {https://doi.org/10.1002/admt.202100016},
	year = {2021}
}
\end{document}